\documentclass{infdis}
\usepackage{hyperref}
\usepackage{cite}

\begin{document}

\title{Socioeconomic bias in influenza surveillance}

\author{Samuel V. Scarpino, James G. Scott, Rosalind M. Eggo, Bruce Clements, Nedialko B. Dimitrov, and Lauren Ancel Meyers\\[4pt]
\affiliation{Northeastern University, Boston, MA, USA; The University of Texas at Austin, Austin, TX, USA; London School of Hygiene and Tropical Medicine, London, UK; Texas Department of State Health Services, Austin, TX, USA; Santa Fe Institute, Santa Fe, New Mexico, USA, ISI Foundation, Turin, Italy}}

\infheader{Scarpino et al.}{Socioeconomic bias in surveillance}

\infabstract{
\abshead{Background}
Individuals in low socioeconomic brackets are considered at-risk for developing influenza-related complications. Adequate influenza surveillance in these at-risk populations is a critical precursor to accurate risk assessments and effective intervention. However, the primary US influenza surveillance system (ILINet) monitors outpatient healthcare providers, which may be largely inaccessible to lower socioeconomic populations. 

\abshead{Methods}
We use a flexible statistical framework for integrating multiple surveillance data sources to evaluate the adequacy of traditional (ILINet) and next generation (BioSense 2.0 and Google Flu Trends) data for forecasting influenza hospitalizations across poverty levels. 

\abshead{Results}
We find that zip codes in the highest poverty quartile are a critical blind-spot for ILINet that the integration of next generation data fails to ameliorate. The models make the best predictions in the most affluent zip codes and the worst predictions in the most impoverished zip codes, regardless of the data sources.   

\abshead{Conclusions}
Here we develop a method to design robust and efficient forecasting systems for influenza hospitalizations.  With these forecasting models, we uncover a key data blindspot, namely that the US public health surveillance data sets perform poorly for the most at-risk communities.  Thus, our study identifies another related socioeconomic inequity.
}

\conflict{Potential conflicts of interest: none reported.}
\corresp{Reprints or correspondence:
Lauren Ancel Meyers, The University of Texas at Austin,
Austin, TX 78712 (laurenmeyers@austin.utexas.edu).}

\firstcolumn{%
As part of a broader national security strategy, the President of the United States created the first \emph{National Strategy for Biosurveillance}, outlining the nation's key strategic goals in disease surveillance\cite{Obama:2012}. As a core component of this strategy, President Obama lists taking ``full advantage of the advanced technologies... that can keep our citizens safe.''  The surveillance systems outlined by the president are targeted at both recurring diseases, such as influenza, and newly emerging infections.  Biosurveillance using advanced technologies may be most important in poor
socioeconomic areas, where influenza burden tends to be
highest~\cite{sloan:etal:2015,tam:etal:2014, placzek2014effect}.\\

This article focuses on the ability of existing influenza surveillance data sets and technologies to predict influenza in at-risk populations, specifically based on the proportion of the population living below the poverty line. Traditional influenza surveillance is based on primary healthcare provider reports, which may be biased towards serving populations with higher socioeconomic status because of the costs and accessibility of healthcare~\cite{nsoesie2015computational, stoto2012effectiveness}.  Next generation data sources provide promise for improving the timeliness and statistical power of surveillance systems.
}

\secondcolumn{%
However, a systematic evaluation of the current surveillance system is needed
to evaluate where it falls short, and whether new data can fill gaps. 
\\

New technologies have fueled a rapid expansion of data sources that can be acquired quickly and inexpensively for public health surveillance. For example, Google Flu Trends used internet search queries of influenza-related terms for surveillance~\cite{ginsberg2009detecting}. The introduction of Google Flu Trends, digital disease surveillance has exploded with efforts on data from search engines (Yuan et al. 2013), crowd-sourced participatory surveillance (e.g., Flu Near You, InfluenzaNet), Twitter (e.g., MappyHealth), Facebook (Boulos et al. 2010; Chew et al. 2010; Lee 2010; Seifter et al. 2010; Broniatowski et al. 2013; Chunara et al. 2013), and Wikipedia access logs (McIver and Brownstein 2014; Generous et al. 2014).
There is evidence that essentially all of these next-generation surveillance data streams correlate to some degree with epidemiological time-series during typical seasonal outbreaks.  
}

\noindent However, there are at least two recent findings worth considering with respect to the these high-tech surveillance systems: 1.) the performance of Google Flu Trends has been unreliable during anomalous influenza outbreaks~\cite{olson2013reassessing, lazer2014parable} and 2:) it is unclear who is responsible for maintaining these systems~\cite{althouse2015enhancing}, especially considering that Google Flu Trends was recently taken offline. \\

Newly upgraded hospital information systems are another promising source of surveillance data.  For example, the United States Centers for Disease Control and Prevention (CDC) launched the BioSense 2.0 program, a set of cooperative agreements between the Department of Veterans Affairs, the Department of Defense, and civilian hospitals from around the country.  Through the cooperative agreements, the BioSense 2.0 program creates a ``data-exchange system that enables its users ... to track health issues as they develop and to share this information quickly''~\cite{BioSense20}.  Whereas BioSense 2.0 provides real-time data on severe cases, the CDC's primary influenza surveillance system, the influenza-like-illness network (ILINet), provides weekly estimates of number of patients presenting with influenza-like-illness symptoms at primary care clinics.  Integrating potentially complementary information from new and traditional systems like BioSense 2.0 and ILINet, along with publicly available internet-source data, like Google Flu Trends, may provide a more timely, comprehensive, and robust picture of disease activity.  To this end, the Defense Threat Reduction Agency has begun a national effort to build the Biosurveillance Ecosystem, an integrated disease surveillance system providing access to diverse data sources and powerful analytics~\cite{DTRAbsve}.  

Here, we analyze the effectiveness of an integrated influenza surveillance system that combines data from BioSense 2.0,  ILINet, hospital discharge records, and historical Google Flu Trends records. At the state and multi-county regional levels, these data sources provide effective situational awareness.  However, we find that they are much more representative of higher socioeconomic sub-populations and perform poorly for the most \emph{at-risk} communities. Thus, the integration of internet and electronic medical records data into surveillance systems may improve timeliness and accuracy, but  fail to remedy a critical blindspot. 

\section{Methods}
\subsection{Short term predictions}

\label{sec:methoddetails}

We used generalized additive models to make short-term predictions of influenza-related hospitalizations in the study populations.  First, we partitioned zip codes into four poverty quartiles. To predict hospitalizations for group $i$, we use the Poisson generalized linear model given by
\begin{align}
    y^{(i)}_{t} \sim \mbox{Poisson}(\lambda^{(i)}_{t}) \; , \quad \log \lambda^{(i)}_{t} = \alpha^{(i)} +  \sum_{k=1}^D h^{(i)}_{k}(x_{k,t}) \, , \label{eqPReg}
\end{align}
where $y^{(i)}_{t}$ is the total number of hospitalizations in group $i$ at
time $t$, $x_{k,t}$ is the $k$th predictor for hospitalizations
at time $t$, $\alpha_i$ is a background hospitalization rate for group $i$,
and $h^{(i)}_{k}(\cdot)$ is some potentially nonlinear function (specific to group $i$) that maps predictors to expected hospitalization counts.  Intuitively, the $x_{k,t}$ scalars capture
all the information used by the surveillance model to predict hospitalizations.   Here $t$ indexes the time of the prediction and $k$ the
particular data source---for example, Google Flu Trends data from two weeks prior.
We fit the $h^{(i)}_{k}(\cdot)$ by expanding each predictor in a third-order B-spline basis with six
degrees of freedom.  To avoid overfitting, we regularized the spline
coefficients using a lasso penalty, with the regularization
parameter chosen by cross-validation.  

Let $y_i = (y_{i1}, \ldots, y_{iN})^T$ be a vector of counts for zip code group $i$ across all weeks.  Let $X$ be an $N \times D$ matrix of surveillance variables used as predictors, where rows are weeks and columns are variables.  We considered one-week-ahead forecasts, thus entry $t$ in $y_i$ corresponds to this week's hospitalization count, while row $t$ of the $X$ matrix (used to forecast $y_{it}$) corresponds to information based on surveillance variables up through week $t-1$ only. Two-week-ahead forecasts were similar, but with the $X$ matrix containing data only through week $t-2$.

We considered four different model variations, each using a different combination of data from BioSense 2.0, ILINet and Google Flu Trends. Each of these three data sources included multiple time series. For example, BioSense 2.0 provided hospitalization counts for each of the six counties in the study area. For each time series included in a model, we added three columns in the $X$ matrix:  the level (actual value of the time series in the trailing week), the slope of that variable (first difference over the trailing two weeks at the time of prediction), and the acceleration (second difference over the trailing three weeks at the time of prediction).  The columns of $X$ corresponded to the predictors in each model, and we modeled the sets of predictors: (i) ILINet alone (15 predictors), (ii) BioSense 2.0 alone (18 predictors), (iii) ILINet+ BioSense 2.0 (33 predictors), and (iv) GFT + ILINet+ BioSense 2.0 (51 predictors).

For each variation, we fitted separate models to each group $i$; these group-level models shared the same predictors, but result in different regression
coefficients from B-spline expansions of each partial response function.  
Overall, we fitted 16 models, one for each combination of zip code group ($i$) and candidate predictor set described.  Given that we had 188 weeks of data and between 15 and 51 predictors per model, we regularized the coefficient estimates in order to avoid over-fitting.  Specifically, we applied a lasso penalty on the coefficient vector $\beta$, by minimizing the objective function
$$
f(\beta) = l(\beta) + \lambda \ p(\beta) \, ,
$$
where $l(\beta)$ is the negative log likelihood arising from the Poisson model, $p(\beta)$ is the lasso penalty function, and $\lambda$ is a scalar that governs the strength of regularization.  We select $\lambda$ for each regression separately using cross validation.  See \cite{mazumder:friedman:hastie:2009} for further details of the model-fitting algorithm.

\section{Predictive performance}
To evaluate the predictive performance of the models, we calculated
out-of-sample RMSE (ORMSE).  Let $\hat{y}_{it}$ be the predicted hospitalization count for group $i$ on week $t$, generated from fitting the model to every data point except week $t$.  The quantity
$$
e_{it} = y_{it} - \hat{y}_{it}
$$
is the out-of-sample prediction error.  We refitted the model 188 times, one for each week that is removed; this is repeated for every group and every combination of surveillance variables.  The ORMSE for a group of zip codes $i$ is given by
$$
\mbox{ORMSE}^{(i)} = \frac{\sqrt{\frac{1}{N} \sum_{t=1}^N e_{it}^2}}{\mathrm{Pop}^{(i)}} \, ,
$$
where $N$ is the number of weeks, and $\mathrm{Pop}^{(i)}$ is the total population of the group.  This can be interpreted as the average predictive error of the model.  The units are hospitalization counts per person.  Although the groups have approximately the same population size, normalizing by the population of the group is essential.  Without normalization, predictions for a large population may appear worse than predictions for a small population, simply because more hospitalizations occur in the larger group.
We corroborated our ORMSE results using a log-likelihood analysis (see Supplement). 

To determine whether performance differences between poverty groups were statistically significant, we ran a permutation test with 10,000 repeats, by randomly assigning zip codes into four equally sized groups, and re-fitting the model to each randomized group, following the original procedure, including cross-validation regularization. We then calculated ORMSE for each group, and also the difference between the best ORMSE and the worst ORMSE among the four groups. 

For each of the four model variants, we (1) used this procedure to generate null distributions of test statistics for each of our four model variants, (2) calculated the difference between the ORMSE measured for the highest poverty quartile and that measure in the lowest poverty quartiles (according to the original grouping), and (3) determined the proportion of the null distribution less than this difference. This proportion was the \textit{P}-value used to determine statistical significance.

\subsection{Data sources}

\label{sec:datasources}

We used the following sources, which contained data primarily from Dallas, Tarrant, Denton, Ellis, Johnson, and Parker counties in Texas, between 2007 and 2012:
\begin{enumerate}

\item{}Weekly BioSense 2.0 data was extracted from an online repository~\cite{BiosenseWeb}.  Data are percent of emergency department (ED) visits for
    Upper Respiratory Infection (URI), based on classification of free-text
    chief complaint entries. Although zip code level data are available, we used county-level aggregates in our analysis.

\item{}ILINet gathers data from thousands of healthcare providers across the USA. Throughout influenza season, participating providers are
    asked to report weekly the number of cases of influenza-like illness
    treated and total number of patients seen, by age group. The case definition requires fever in excess of $100^{\circ}$F with a cough
    and/or a sore throat without another known cause. The Texas Department of State Health Services (DSHS) provided weekly ILINet records
from $2007-2012$, aggregated by county. 

\item{}Google Flu Trends estimated the
    number of ILI patients per $100,000$ people based on the daily number of
    Google search terms associated with signs, symptoms, and treatment for
    acute respiratory infections.  Although Google Flu Trends is no longer active, past data are available for download from
    Google.org and has been shown to reliably estimate
    seasonal influenza activity~\cite{ginsberg2009detecting,Valdivia:2010}, but be unreliable for
    the $2009$ H1N1 pandemic~\cite{Wilson:2009}. We considered six different Google Flu Trends time series, corresponding to the state of Texas and five cities in the Dallas-Fort Worth area (Fort Worth, Irving, Plano, Addison
    and Dallas.) 

\end{enumerate}

The surveillance models predict hospitalizations by zip code.  We obtained hospital discharge records from Texas
Health Care Information Collection (THCIC), filtered for influenza-related principal diagnostic codes of
ICD-9 487.*, which includes 487.0 (with
pneumonia), 487.1 (with other respiratory manifestations) and 487.8 (with other
manifestations). The data are aggregated into weeks and by patient zip code. 

Supplemental Figure 2 presents aggregate counts from the BioSense 2.0, ILINet and
hospitalization data used in the study for the  Dallas-Fort Worth region.
We grouped zip codes into poverty quartiles, based on the percentage of the
population living in poverty reported in the 2011 American
Community Survey~\cite{acs2011}. We estimated age distributions within zip codes from the 2011 American Community Survey and the 2010 Census.

\subsection{Institutional Review Board Approval}
The Texas Department of State Health Services Institutional Review Board \#1 approved this project.  The associated reference number is IRB\# 12-051.  An informed consent waiver was approved by the IRB.

\section{Results}
\begin{figure*}[!t]
\centering{\includegraphics[width=0.45\textwidth]{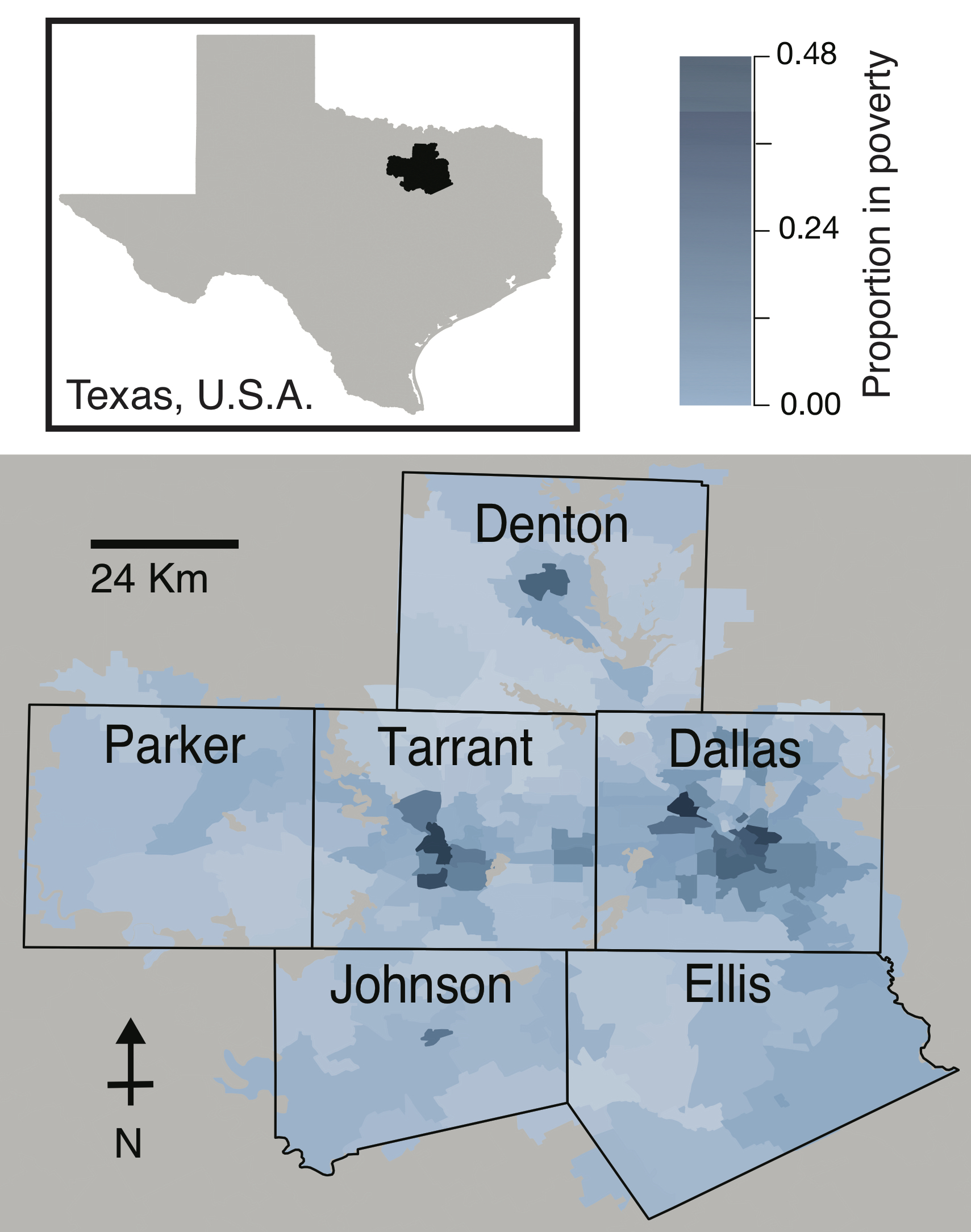}}
\caption{The proportion of each zip code living below the federally defined poverty level in the six-county study region.}
\label{figMap}
\end{figure*}

We evaluated the performance of BioSense 2.0, Google Flu Trends and ILINet data sources, with respect to short-term predictions of influenza-related hospitalizations in the six-county region surrounding the Dallas-Fort Worth metropolitan area (Figure \ref{figMap}). This region included 305 zip codes and all of the emergency departments reporting to the Texas BioSense 2.0 system during the five-year study period (2007-2012).

\subsection*{Influenza burden by poverty level and age}
\begin{figure*}[!t]
\centering{\includegraphics[width=\textwidth]{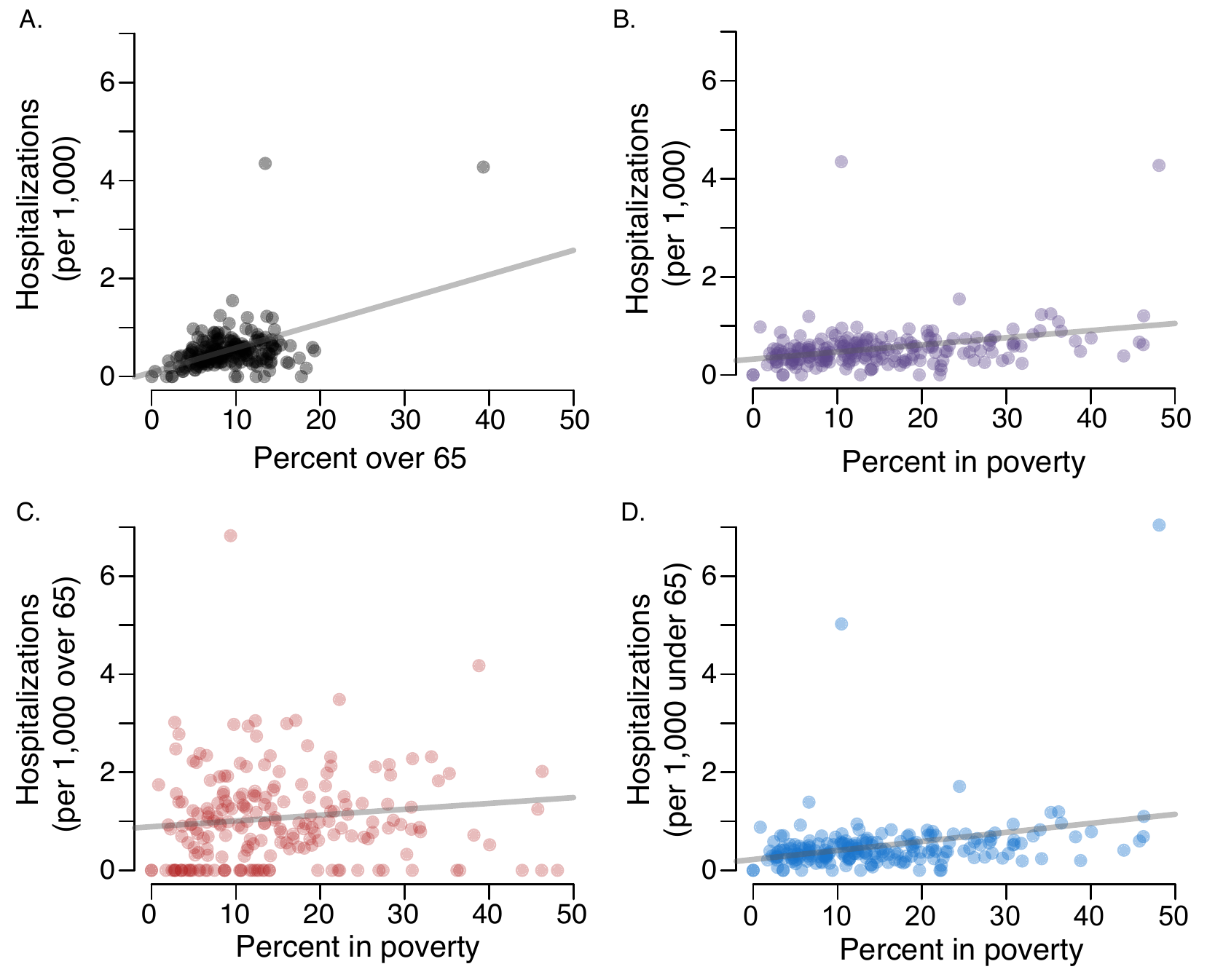}}
\caption{Relationship between age, poverty level, and influenza hospitalizations across 305 zip codes from 2007 to 2012. Demographic data are based on 2010 Census. (A) Influenza hospitalizations increase with the size of the over 65 population (p \textless .001). (B) Influenza hospitalizations increase with the percent of the population under the federal poverty level (p \textless .001). 	(C) Influenza hospitalizations in over 65 year olds does not significantly increase with poverty (p = .11). (D)  Influenza hospitalizations in under 65 year olds does  significantly increase with poverty (p \textless .001). }
\label{figPoverty}
\end{figure*}

We estimated the influenza hospitalization
rate per 1,000 people in each zip code. Throughout the region, we find that influenza 
hospitalization rates exhibit a significant positive correlation with both poverty level and the proportion of the 2010 census population over age 65 (Figure \ref{figPoverty}), consistent with recent literature ~\cite{Thompson2004,sloan:etal:2015,tam:etal:2014}. 
Stratifying by age group, poverty and influenza burden are significantly correlated in the under age 65 but not the over age 65 populations (2011
American Community Survey estimates) (\textit{p}\textless.001). We established this result with a multivariate regression of hospitalization rate at the zip-code level with the proportion of the zip code living below the poverty line and the proportion of the zip code over 65  (Table~\ref{tab1}).

\begin{table}[ht]
\begin{small}
\begin{center}
\begin{tabular}{r c c c c}
  \hline
 & Estimate & Std. Error & t value & Pr($>$$|$t$|$) \\ 
  \hline
(Intercept) & 0.3656 & 0.0978 & 3.74 & 0.0002 \\ 
  Proportion over 65 & -0.0001 & 0.0094 & -0.01 & 0.9911 \\ 
  Proportion in poverty & -0.0096 & 0.0043 & -2.25 & 0.0258 \\ 
  Interaction & 0.0022 & 0.0003 & 6.26 & 0.0000 \\ 
   \hline
\end{tabular}
\caption{A multivariate regression of hospitalization rate at the zip-code level with the proportion of the zip code living below the poverty line and the proportion of the zip code over 65.}
\label{tab1}
\end{center}
\end{small}
\end{table}

\subsection*{Data quality by poverty quartile}
\begin{figure*}[!t]
\centering{\includegraphics[width=\textwidth]{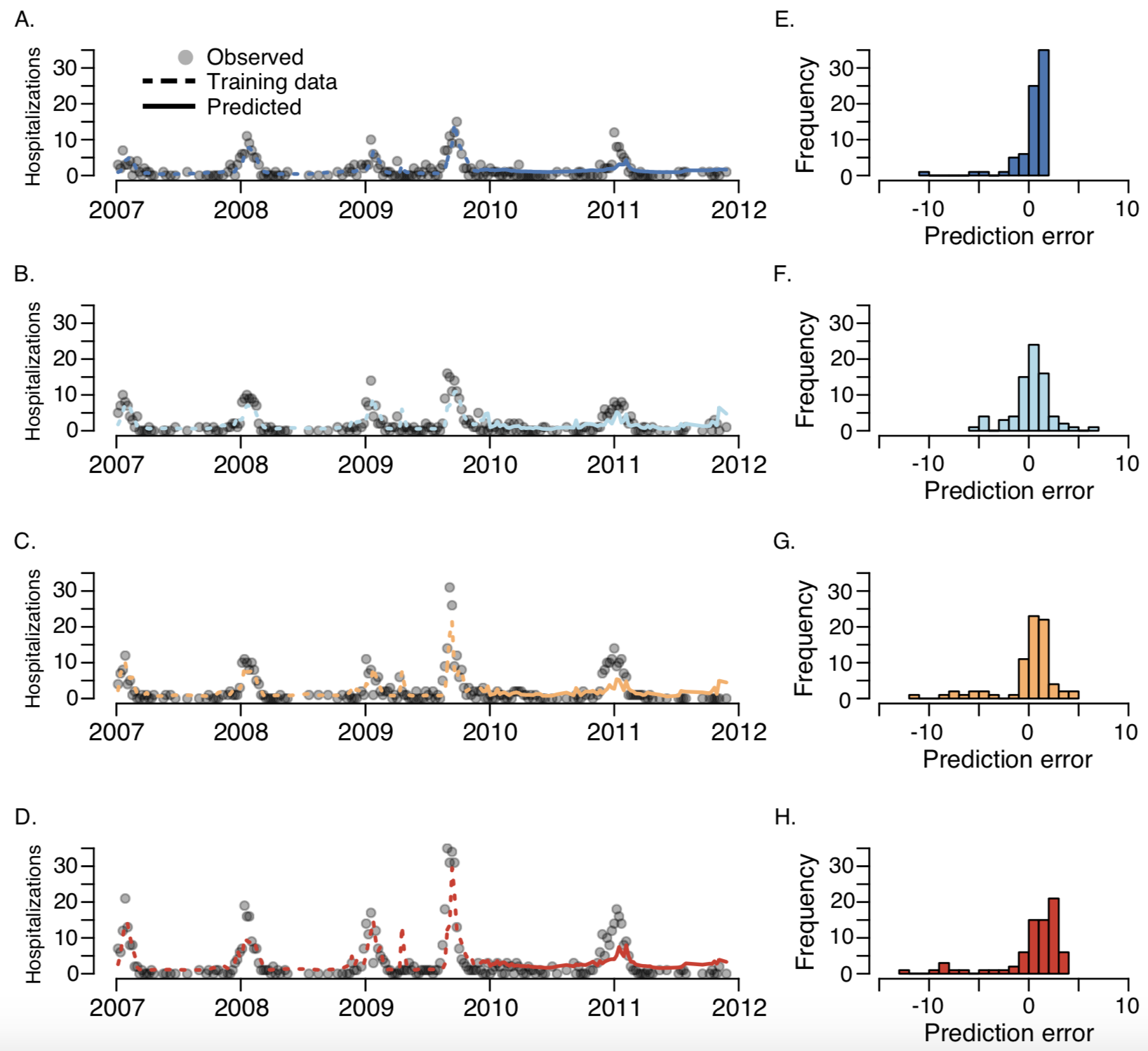}}
\caption{Comparison between one-week ahead model predictions and the total number of weekly observed influenza hospitalizations for each of the four poverty quartiles (A) Upper quartile, (B) Upper-middle quartile, (C) Lower-middle quartile, (D) Lowest quartile and the distribution of out-of-sample prediction errors (observed - predicted) for the (E) Upper quartile, (F) Upper-middle quartile, (G) Lower-middle quartile, and (H) Lowest quartile.  The model was trained on the first 60\% of the data (dashed lines) and evaluated on the remaining 40\% of the data (solid lines).  Qualitatively similar results were obtained with n-fold (leave-one-out) cross-validation and 80\slash 20 training\slash testing, see Supplement.  Across all four quartiles, the model was unbiased according to a re-sampling test on the residuals, see Supplement.}
\label{figFit}
\end{figure*}

We classified zip codes into poverty quartiles based on the proportion of the population living below the federally defined poverty line and fitted separate generalized additive forecasting models to the data in each of the quartiles. In comparisons between model predictions and hospitalization data, we find that the data become less informative as the poverty level increases (Figure \ref{figFit} and Table~\ref{tab2}). The models make the best predictions in the most affluent 25\% of zip codes---with poverty levels between 0\% and 7.5\%--- and the worst predictions in the most impoverished 25\% of zip codes---those with poverty levels between 21.2\% and 48.1\%, regardless of the data sources included as predictors. The differences in prediction errors between the upper and lower poverty quartiles are statistically significant ($\textit{P}<0.0001$, bootstrap analysis).  This trend is confirmed by out-of-sample Poisson log-likelihoods, along with negative binomial and least squares regressions (also evaluated on out-of-sample data, see Supplement). 

\begin{table}[ht]
\begin{small}
\begin{center}
\resizebox{\columnwidth}{!}{%
\begin{tabular}{r c c c c c}
\hline
 Surveillance Data Sources & 1st quartile & 2nd quartile & 3rd quartile & 4th quartile & Combined\\ 
  \hline
ILI & 1.69 & 2.41 & 2.29 & 5.12 & 2.22 \\ 
  Biosense & 1.55 & 1.95 & 2.51 & 2.60 & 2.01 \\ 
  GFT & 1.38 & 1.34 & 2.16 & 2.68 & 1.74 \\ 
  ILI + Biosense & 1.46 & 1.68 & 2.30 & 3.81 & 1.94 \\ 
  ILI + GFT & 1.42 & 1.35 & 2.17 & 2.75 & 1.74 \\ 
  Biosense + GFT & 1.44 & 1.58 & 2.11 & 2.64 & 1.79 \\ 
  ILI + Biosense + GFT & 1.44 & 1.53 & 2.12 & 2.64 & 1.72 \\ 
\end{tabular}
}
\caption{Out-of-sample (60\slash 40 training\slash testing) root mean-squared error (ORMSE) using a Poisson generalized additive model. Values are normalized by the population size of each zip code quartile and then multiplied by $10^6$ to obtain ORMSE per one million residents. The rightmost column gives aggregate ORMSE across all Texas zip codes.}
\label{tab2}
\end{center}
\end{small}
\end{table}

We further evaluated the models fit by training on the first 60\% and 80\% of the time series, and testing on the held-out 40\% and 20\% respectively, and obtained qualitatively similar differences in performance across poverty quartiles (see Supplement).  Finally, we use a simulation to show that the disparity in coverage cannot be explained by higher case hospitalization rates in more impoverished populations (see Supplement). All else being equal, a higher hospitalization rate should provide additional data and thus statistical power for epidemiological predictions, while a lower hospitalization rate or reduced sampling by surveillance data sources might impair predictions.  

\begin{figure*}[!t]
\centering{\includegraphics[width=\textwidth]{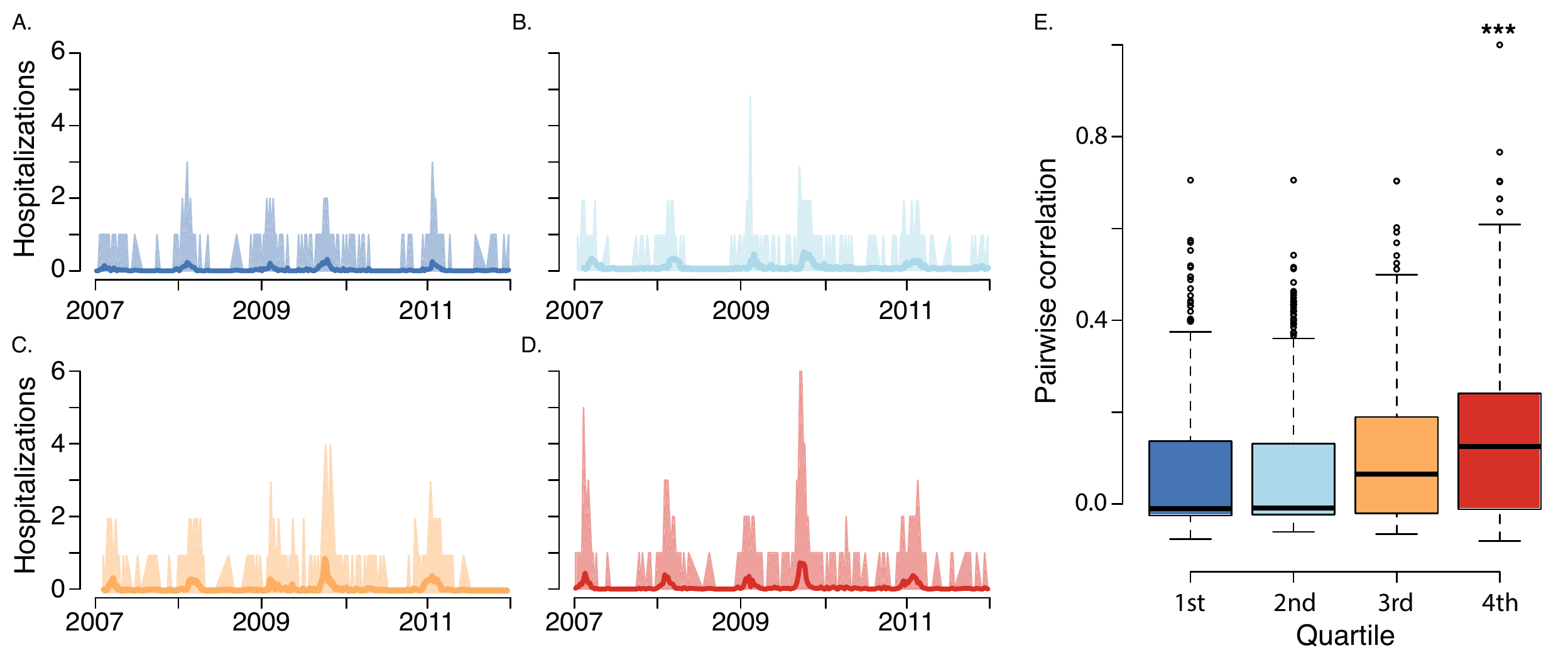}}
\caption{Influenza synchrony among zip codes within each  poverty quantile.  (A-D) Range of influenza activity (shades) around mean (solid line) at the zip code level, in lowest poverty to highest poverty quartiles, respectively. (E) Boxplots of correlation coefficient among pairs of zip codes within each quartile. The poorest quartile exhibited the greatest synchrony.  To test for significance, we randomly assigned zip codes to poverty group 5,000 times and repeated the analysis.  The observed mean and median correlation where higher than those in all 5,000 (mean) and all be but 2 of the 5,000 (median).  We confirmed these results using a principle component analysis, see Supplement.}
\label{figCoherence}
\end{figure*}

\subsection*{Synchrony within poverty quartiles}
We tested the hypothesis that the lower performance observed in the most disadvantaged quartile resulted from greater asynchrony in influenza hospitalization rates among the constituent zip codes. We define asynchrony as the average pair-wise correlation between zip codes.  Based on data visualization and pairwise correlation analyses among zip codes, we failed to find evidence in support of this hypothesis (Figure~\ref{figCoherence}). In fact, influenza hospitalization patterns exhibited significantly more similarity within the lowest poverty quartile than within the less impoverished quartiles.  We confirmed these results using a principal-component analysis of zip code level hospitalizations (see Supplement).

\section{Discussion}
Populations with lower socioeconomic status have higher hospitalization rates across a range of diseases~\cite{billings1993, placzek2014effect}, caused in part by reduced access to healthcare~\cite{shi1999income}.  Our analysis suggests a similar disparity in the accuracy of public health outbreak surveillance.

Specifically, a combination of clinical symptom reports, internet searches, and electronic emergency room data can predict week-ahead inpatient influenza hospitalizations more reliably in higher socioeconomic than in lower socioeconomic populations.  Given this performance discrepancy, we were surprised to find that high poverty zip codes exhibit much more synchronous influenza hospitalization patterns than low poverty zip codes. Thus, the failure likely stems from data bias or under-sampling of at-risk populations. We speculate that  Google Flu Trends (which tallies the number of influenza related Google searches) and ILINet (which collects data from volunteer outpatient clinics) provide low coverage of at-risk populations~\cite{stoto2012effectiveness, nsoesie2015computational}, while BioSense 2.0 may be biased by an excess in non-emergency visits to emergency rooms among uninsured and Medicaid recipients~\cite{gandhi2014trends}.

Over 100 years of epidemiological study demonstrates a consistent, negative association between health and economic prosperity~\cite{farmer2001infections, marmot2005social}.  In many settings, lower socioeconomic status has been linked to both reduced access to healthcare and increased burden of both infectious and chronic diseases~\cite{shi1999income, liao2004reach, mensah2005state, kandula2007differences}.  For example, the REACH 2010 surveillance program in the U.S.A. found that, ``More minorities reported being in fair or poor health, but they did not see a doctor because of the cost."~\cite{ liao2004reach}. In this vein, we found positive correlation between poverty and influenza hospitalization rates in study populations under age 65, which is consistent with a three-fold excess in pediatric influenza-related hospitalizations estimated for a Connecticut at-risk community~\cite{yousey2011neighborhood}. However, it is unknown which of many possible factors---including differences in sanitation, crowding, vaccine coverage, or prevalence of underlying conditions---are driving this disparity.

Our study identifies another related socioeconomic inequity---a reduced capability to detect and monitor outbreaks in at-risk populations---which impedes effective public health interventions. A analogous surveillance gap has been identified for cancer~\cite{glaser2005cancer}. Ironically, surveillance systems seem to neglect communities most in need of intervention. New methods for designing and optimizing disease data collection have focused on state-level coverage~\cite{polgreen2009optimizing, scarpino2012optimizing} or assumed that risk was evenly spread across well-mixed populations~\cite{ pelat2014optimizing}, but could be adapted to identify data sources that remedy critical gaps or biases.

A growing community of researchers and practitioners across public health, medicine, science, military, and non-governmental organizations are developing and deploying technology-enabled surveillance systems~\cite{althouse2015enhancing}.  Many of these efforts focused on improving the timeliness and accuracy of bioevent detection, situational awareness, and forecasting. However, our results suggest a different, and arguably more important priority: improving coverage in at-risk populations. Gaps in both traditional and early next generation surveillance systems compound health disparities in populations with reduced access to healthcare or higher rates of severe disease.  Thus, as surveillance systems are upgraded and expanded to incorporate  novel data sources, particular attention should be paid to improving equity, in addition to other performance goals~\cite{lazer2014parable, althouse2015enhancing}.

We recognize several important limitations of our study.  First, our analysis was restricted to the Dallas-Fort Worth region from which we obtained BioSense 2.0 data, and may not generalize to the rest of the USA. Second, since we could not access BioSense 2.0 with influenza diagnoses, we used upper respiratory infections data as a proxy. We expect that influenza-specific BioSense 2.0 records would generally improve one-week-ahead predictions, but may or may not close the surveillance poverty gap. Third, we did not consider many other data sets, some of which might provide more representative coverage of at-risk populations, including public health laboratory data, school absenteeism records, or other internet-sourced or social media data. Fourth, because we used a lasso penalty to regularize the regression coefficients--implying that the number of degrees of freedom does not necessarily increase with the number of predictors--we could not apply standard model selection methods, such as Akaike Information Criteria, to compare the performance across models (rows of Table~\ref{tab2}).  Although BioSense 2.0 yields slightly higher performance scores across all poverty quartiles, we leave a definitive comparison among different combinations of surveillance data sources for future study.  Fifth, we did not have individual-level patient socioeconomic and/or zip code information from ILINet, BioSense 2.0, and GFT, and thus we were unable to assess directly whether lower socioeconomic groups are underrepresented.  However, prior studies suggest that lower socioeconomic groups use the internet less frequently than higher socioeconomic groups, and that disease-related signals derived from Internet-search data poorly reflect incidence in lower socioeconomic communities~\cite{richiardi2014, silver2014}. Researchers with access to individual-level BioSense 2.0 and GFT data could test our hypothesis, and perhaps develop methods for subsampling the data to improve predictive performance in low income areas.    Finally, the Texas inpatient hospitalization data did not indicate whether patients were admitted through an emergency department.  Therefore, we were unable to determine whether visitation rate to emergency departments for influenza varied by socioeconomic status.  We note that the majority of inpatient hospitalizations in the US are not preceded by an emergency department visit~\cite{schuur2012}. 

\section{Conclusions}
We introduce a robust and flexible method for improving and bench marking situational awareness. Our method offers a general statistical model for short-term prediction, that can systematically integrate diverse data sources, including traditional surveillance data, electronic medical records and internet-source digital data. We used the method to construct a surveillance system that made one-week-ahead predictions of influenza hospitalizations from real-time BioSense 2.0, Google Flu Trends and ILINet data. While overall performance was reasonable, we discovered a critical data blindspot in Dallas-Fort Worth's most at-risk populations. This surveillance design framework can be readily applied to evaluate and integrate new data sources that address this challenge.   

\section{Conflicts}
All authors declare that no conflicts of interest exist.

\begin{funding}
This work was supported by the National Institutes of Health, the Omidyar Foundation, and the Santa Fe Institute.
\end{funding}

\begin{acks}
The authors would like to acknowledge funding support from MIDAS U01 GM087719 to LAM.  SVS acknowledges funding from the Omidyar Group and the Santa Fe Institute.
\end{acks}


\end{document}